# NeuroFlake: A Neuro-Symbolic LLM Framework For Flaky Test Classification

Khondaker Tasnia Hoque*
bsse1205@iit.du.ac.bd
Institute of Information Technology, University of Dhaka
Dhaka, Bangladesh

Toukir Ahmed
toukir@iit.du.ac.bd
Institute of Information Technology, University of Dhaka
Dhaka, Bangladesh

## Abstract

Flaky tests, which exhibit non-deterministic pass/fail behavior for the same version of code, pose significant challenges to reliable regression testing. While large language models (LLMs) promise for automated flaky test classification, they often fail to comprehend the actual logic behind test flakiness, instead overfitting to superficial textual artifacts (e.g., specific variable names). This semantic fragility leads to poor generalization on real-world imbalance dataset and vulnerability to perturbations. In this paper, we introduce NeuroFlake, a novel neuro-Symbolic framework for classifying flaky tests on highly imbalanced, real-world datasets (FlakeBench). Unlike prior approaches that rely on brittle manual rule and black box learning, NeuroFlake integrates a Discriminative Token Mining (DTM) module to automate the discovery of high-fidelity, statistically significant source code tokens (e.g., specific concurrency primitives or async waits). By injecting these strong latent signals directly into LLM's attention mechanism, we bridge the gap between neural intuition and symbolic precision. Our experiments demonstrate that neuro-symbolic fusion significantly improves classification performance by leveraging classification F1-score to 69.34% while prior state-of-art shows best F1-score 65.79%. However, we rigorously evaluate NeuroFlake's robustness through adversarial stress testing, introducing semantic preserving augmentations (e.g., dead code injection, variable renaming). While baseline models exhibit performance degradation of 8-18 percentage points (pp) on perturbed tests, NeuroFlake maintains performance stability on unseen augmentations dropping only 4-7 pp.

## 1 Introduction

Continuous Integration (CI) and regression testing are the backbones of modern software reliability, ensuring code correctness throughout the development lifecycle [2, 45]. Modern CI pipelines rely heavily on automated test execution to provide rapid feedback to developers. However, the effectiveness of this process is frequently undermined by flaky tests—tests that non-deterministically pass or fail on the same version of code without any modification [27]. In other words, without any changes to the source code or test environment, a flaky test might pass in one CI build but fail in the next, making it completely unreliable. Flakiness misleads developers into investigating non-existent bugs, causing a significant waste of debugging resources and eroding trust in the testing infrastructure [34].

To mitigate this issue, automatic flaky test detection has emerged as an important research direction. However, simply detecting that a test is flaky is insufficient to resolve the issue. Flakiness stems from diverse root causes—such as concurrency race conditions [42], asynchronous waits [41], or test-order dependencies [22]—each requiring a fundamentally different mitigation strategy. For instance, increasing a timeout will not fix a race condition, and adding a lock will not resolve a state-pollution issue. Consequently, treating all flaky tests as a single monolithic category forces developers to manually inspect execution logs to diagnose the root cause, a prohibitively expensive process in large-scale CI environments. Therefore, fine-grained automated classification is not merely a diagnostic aid; it is a prerequisite for (1) Extracting flaky tests from CI pipelines to prevent false alarms [20], (2) Introducing repair strategy according to flakiness [7, 21, 25], and (3) Helping developers understand the root cause of the failure [17].

Early approaches to flaky test detection relied on execution-based techniques such as repeated test runs and controlled scheduling [4, 18], which are often expensive and impractical at scale. Consequently, research has shifted toward static automated prediction using Machine Learning (ML) [24, 35, 43] and Deep Learning (DL) [1]. However, these traditional models often rely on hand-crafted features or shallow statistical representations, failing to capture the complex, context-dependent semantic logic required to distinguish subtle flakiness categories. Recent advances in Large Language Models (LLMs) [10] have demonstrated state-of-the-art performance in flaky test classification. Large Language Models (LLMs) like CodeBERT [38] and GraphCodeBERT [13] are fine tuned to classify tests according to their flakiness by treating source code as natural language. Approaches such as Flakify [11] and FlakyCat [1] leverage pre-trained code models to classify flaky tests depending on source code tokens.

However, despite their high accuracy on standard benchmarks, these neural approaches face two critical limitations. First, as highlighted by [40], LLM-based classifiers suffer from shortcut learning [46]: they often memorize superficial syntactical cues (e.g., specific variable names like waitTime) rather than understanding the underlying semantics of flakiness. Second, prior works [1, 3, 11] often assume balanced data distributions. In contrast, real-world datasets like FlakeBench [40] are severely imbalanced, where specific flaky categories (e.g., Concurrency or Test Order Dependency) are rare. Standard models biased toward the majority class fail to detect these rare but critical categories, rendering them ineffective in practice.

These limitations indicate that effective flaky test classification requires models that (1) go beyond token-level cues to capture deeper semantic signals and (2) remain robust under severe class imbalance and adversarial perturbations. Addressing these challenges remains an open problem in existing LLM-based approaches. Fundamental questions arise regarding the design of reliable flaky test classification model. Specifically, we investigate whether a neuro-symbolic framework can transcend superficial token matching to grasp the true semantics of flakiness. Furthermore, we examine whether the integration of symbolic reasoning with neural representations improves performance on rare, imbalanced categories, and assess whether this hybrid approach yields greater robustness against adversarial noise.

To address these challenges, we introduce **NeuroFlake**, a comprehensive neuro-symbolic framework designed to robustly classify flaky tests by overcoming the dual hurdles of shortcut learning and data imbalance. While Chen and Jabbarvand [7] leveraged neuro-symbolic power of LLMs to repair order-dependent flaky tests, no existing method uses neuro-symbolic approach to classify flaky tests from imbalanced dataset, preserving the robustness against token level perturbations [5, 6]. NeuroFlake employs a novel dual-strategy architecture. First, it integrates a Neuro-Symbolic channel [7, 15] that augments the neural embeddings of GraphCodeBERT [13] with a vector of symbolic features extracted via statistical analysis (Discriminative Token Mining) [16, 23]. To handle severe imbalance, we implement Effective Number of Samples (ENS) weighting [8], which optimizes the loss function for rare categories more effectively than based weighting [40]. Second, to enforce semantic understanding, we introduce a Smart Adversarial Augmentation pipeline. By injecting semantics-preserving traps—such as dead code or variable renaming—during training, this force the model to ignore deceptive shortcuts and focus on the structural logic of the test code. Later, We evaluate robustness by stress test on unseen, diagnostic perturbations. Our evaluation on the real-world FlakeBench dataset demonstrates that NeuroFlake significantly outperforms existing baselines. The framework achieves a macro F1-score of 69.34%, a substantial improvement over the prior state-of-the-art of 65.79% [40]. Crucially, stress testing reveals that while baseline models suffer performance drops of more than 18% under perturbation, NeuroFlake maintains robustness with minimal degradation. The implications of these results suggest that hybrid neuro-symbolic architectures are essential for reliable code analysis in imbalanced environments.

In summary, this paper makes the following contributions:

- **Dual-Channel Architecture:** We propose a neuro-symbolic design that combines LLM embeddings with adaptive symbolic features to improve semantic understanding.
- **Robust Training Strategy:** We introduce a Smart Adversarial Augmentation strategy that trains the model to resist perturbations, mitigating shortcut learning [40].
- **Comprehensive Evaluation:** We conduct an extensive evaluation using unseen diagnostic perturbations, demonstrating that NeuroFlake outperforms existing baselines in both classification performance and robustness.

To facilitate reproducibility and future research, the implementation and dataset used in this study are available.[1]

## 2 Related Work

Research into test flakiness began with empirical studies aimed at characterizing its diverse root causes [34]. Luo et al. [27] presented the first empirical study of flaky tests, identifying their causes. Flakiness is not a monolithic defect; rather, it manifests through distinct semantic categories that require specific mitigation strategies. Asynchronous Wait failures occur when a test assumes an operation is complete before the asynchronous task (e.g., a thread or callback) has actually finished [36, 37]. Concurrency issues stem from race conditions or deadlocks triggered strictly under specific thread schedules [19, 42]. Order Dependent (OD) tests fail based on the execution sequence of the suite, typically due to shared state pollution in static fields or databases [14, 39]. Other categories include Time-Dependent failures linked to system time or rigid timeouts, and Unordered Collection bugs where code assumes a specific order in non-deterministic data structures [27, 34]. Since manual identification of these root causes is prohibitively expensive, the field has prioritized automated classification.

The automated classification of flaky tests has evolved into two distinct streams: traditional Machine Learning (ML) and Deep Learning (DL) approaches. Early detection techniques relied on execution-based rerunning strategies [18], which incurred high computational overhead. Consequently, researchers shifted to Machine Learning (ML). ML-based tools like FlakeFlagger [3], Flakat [24] and FlakyRank [43] utilized handcrafted features (e.g., keyword counts, code metrics) to predict flakiness. However, these traditional ML classifiers rely on manual feature engineering, which fails to capture the subtle, data-dependent logic of rare flakiness types. Recently, the field has transitioned to Large Language Models (LLMs) [28, 33, 44], which treat test code as natural language. Approaches like Flakify [11] and FlakyCat [1] leverage these pre-trained models to achieve state-of-the-art accuracy by learning contextual representations from raw source code. While these LLM-based methods outperform ML based classifiers, they face significant limitations when applied to real-world environments. These models suffer from shortcut learning [46] -memorizing syntactic cues (e.g., variable names like sleep) rather than understanding the actual defect logic. Additionally, they are predominantly trained on balanced datasets that do not reflect reality. Rahman et al. [40] highlighted these limitations by introducing FlakeBench, a realistic, highly imbalanced benchmark that better reflect. When FlakyCat, Flakify are tested on FlakeBench [40] dataset, the performance drops significantly, as they bias heavily toward majority classes. Furthermore, Flakylens introduced by Rahman et al. [40] outperforms both Flakify and FlakyCat on Flakebench dataset. But Flakylens is sensitive to token level perturbation.

[1] Anonymous Figshare link: https://figshare.com/s/f981ebdaa8082af9974c

Hence, no prior work jointly addresses multi-class flaky test classification, severe real-world imbalance, and robustness against syntactic perturbations, motivating the need for a robust and semantically-aware classification framework. In parallel, neuro-symbolic approaches have shown promise in improving robustness and interpretability in AI systems [15] such as Chen and Jabbarvand [7] applied neuro-symbolic reasoning to flaky test repair, it is necessary to investigate whether neuro-symbolic fusion can deliver similar benefits for flaky test classification tasks.

# 3 Methodology

We propose NeuroFlake, a neuro-symbolic framework designed to classify flaky tests within highly imbalanced, real-world datasets. NeuroFlake operates as a Dual-Channel Inference Engine, illustrated in Figure 1. The pipeline processes a given test case $T$ through two concurrent streams:

(1) **The Adaptive Symbolic Channel:** Rather than relying on brittle, hard-coded rules (e.g., Regex for `thread.sleep(1000)`), this channel utilizes Discriminative Token Mining (DTM), which statistically isolates discriminative tokens from the training corpus.
(2) **The Semantic Neural Channel:** Simultaneously, the source code is processed by GraphCodeBERT, a pre-trained transformer aware of Data Flow Graphs (DFG). This channel captures variable dependency chains and control flow logic that static keywords miss.

The workflow consists of four main stages:

- **Symbolic Indicator Mining:** Discriminative symbolic indicators are automatically mined from the training corpus using *Discriminative Token Mining (DTM)*, producing compact, category-aware symbolic feature vectors.
- **Neural Code Encoding:** Test code is encoded using GraphCodeBERT, which captures semantic structure through data-flow–aware representations.
- **Late Fusion and Optimization:** Symbolic features are projected into a latent space and late-fused with neural code embeddings only at the final inference stage. The unified representation is optimized using an *Imbalance-Aware Training Strategy* combining Effective Number of Samples (ENS) weighting and focal loss.
- **Robustness Training and Evaluation:** To mitigate shortcut learning [40, 46] and improve robustness, NeuroFlake is trained with semantic-preserving adversarial augmentations (e.g., variable renaming and dead-code insertion) and evaluated using unseen perturbation-based stress tests.

The following subsections describe the components of NeuroFlake in detail.

## 3.1 Discriminative Token Mining (DTM)

Unlike static rule-based systems that rely on hard-coded heuristics (e.g., looking only for `Thread.sleep` for detecting asynchronous flakiness), we introduce a data-driven process called **Discriminative Token Mining (DTM)** to automatically discover the "vocabulary of flakiness". This proposed mining process is formalized as a statistical hypothesis testing problem, detailed in Algorithm 1.

**Algorithm 1** Discriminative Token Mining (DTM)

**Require:** Training corpus $\mathcal{D}$, category set $C$, top-$K$ limit $K$
**Ensure:** Symbolic vocabulary $\mathcal{V}_{sym}$
1: $\mathcal{V}_{sym} \leftarrow \emptyset$
2: $M \leftarrow$ TF-IDF_Vectorize($\mathcal{D}$, stopwords = Java keywords)
3: **for each** category $c \in C$ **do**
4: $\quad \mathcal{D}_{pos} \leftarrow \{d \in \mathcal{D} \mid \text{label}(d) = c\}$
5: $\quad \mathcal{D}_{neg} \leftarrow \{d \in \mathcal{D} \mid \text{label}(d) \neq c\}$
6: $\quad Scores \leftarrow \emptyset$
7: $\quad$ **for each** token $t \in M$ **do**
8: $\quad\quad O \leftarrow \text{ObservedFreq}(t, \mathcal{D}_{pos})$
9: $\quad\quad E \leftarrow \text{ExpectedFreq}(t, \mathcal{D}) \times P(c)$
10: $\quad\quad \chi^2(t, c) \leftarrow \frac{(O-E)^2}{E}$
11: $\quad\quad Scores \leftarrow Scores \cup \{(t, \chi^2)\}$
12: $\quad$ **end for**
13: $\quad TopTokens \leftarrow \text{SelectTopK}(Scores, K)$
14: $\quad \mathcal{V}_{sym} \leftarrow \mathcal{V}_{sym} \cup TopTokens$
15: **end for**
16: **return** $\mathcal{V}_{sym}$

Given the training corpus $\mathcal{D}$, we first vectorize the source code using TF-IDF, filtering out standard Java keywords to isolate domain-specific identifiers. For each flakiness category $C_k$, we perform a Chi-Square ($\chi^2$) Test of Independence [29]. We calculate the divergence between the observed frequency ($O$) of a candidate token $t$ in category $C_k$ versus its expected frequency ($E$) under the null hypothesis that the occurrence of $t$ is independent of $C_k$:

$$\chi^2(t, C_k) = \sum \frac{(O_{ij} - E_{ij})^2}{E_{ij}} \tag{1}$$

Tokens with $p < 0.05$ (the standard statistical threshold to reject the null hypothesis [29])are selected as candidate features. To strictly prevent overfitting, we apply a *Cross-Project Consistency Filter*: a token is retained in the symbolic vocabulary $\mathcal{V}_{sym}$ only if it appears in at least $N = 3$ unique projects within the training set. This constraint ensures that $\mathcal{V}_{sym}$ captures generalizable software engineering concepts (e.g., `socket`, `bind`) rather than project-specific noise.

Figure 2 shows the top 10 predictor tokens based on chi-square ($\chi^2$) score for asynchronous wait flakiness. The horizontal axis represents the $\chi^2$ scores of the mined tokens, while the vertical axis represents the mined token list for async wait category. After applying cross-project filtering, we retain tokens such as `thread sleep`, `sleep(has_sleep)`, `thread(has_thread)`, `timeunit seconds`, and `await` as the most important discriminative tokens for async wait flakiness. The same mining process is applied to the remaining flakiness categories also.

Figure 3 illustrates the final distribution of final mined token groups in each flaky test category. The horizontal axis represents token groups(similar types of tokens are clustered in a single group) and vertical axis represents the flaky test categories.

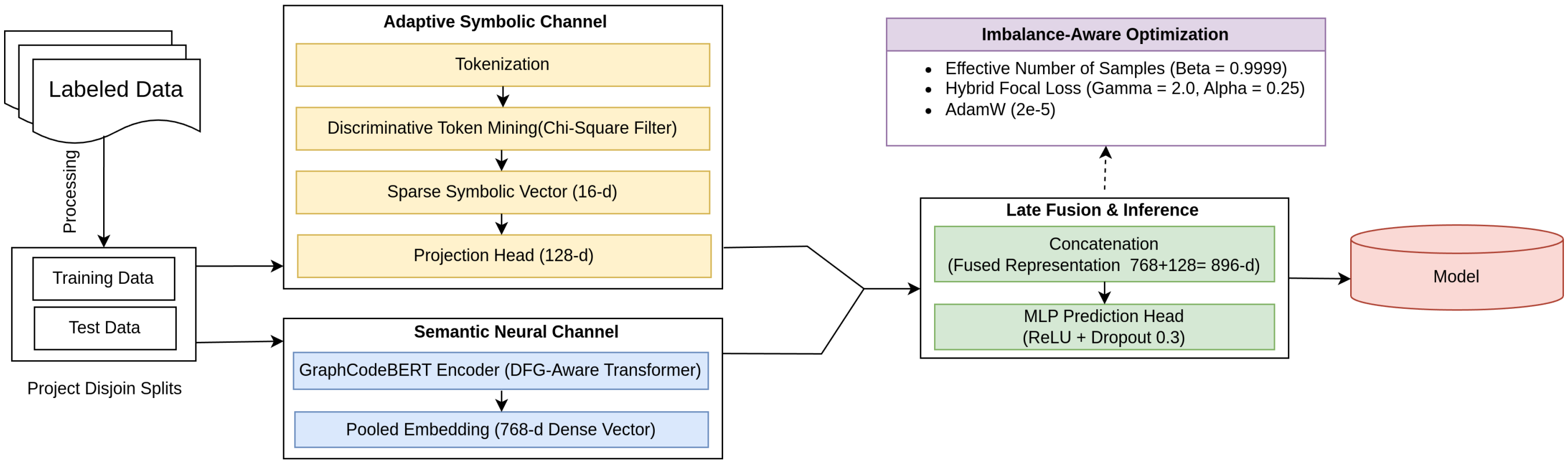

**Figure 1: NeuroFlake Architecture: Dual-Channel Inference Engine.**

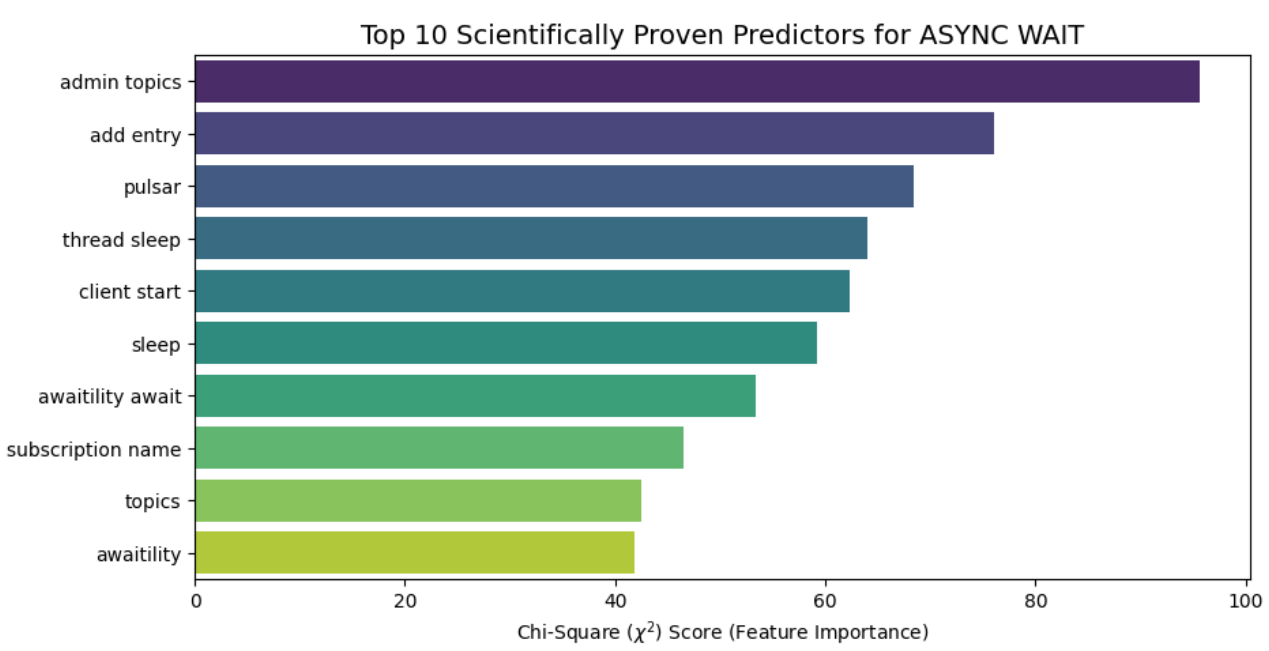

**Figure 2: Top 10 predictor tokens for asynchronous wait flakiness ranked by $\chi^2$ score.**

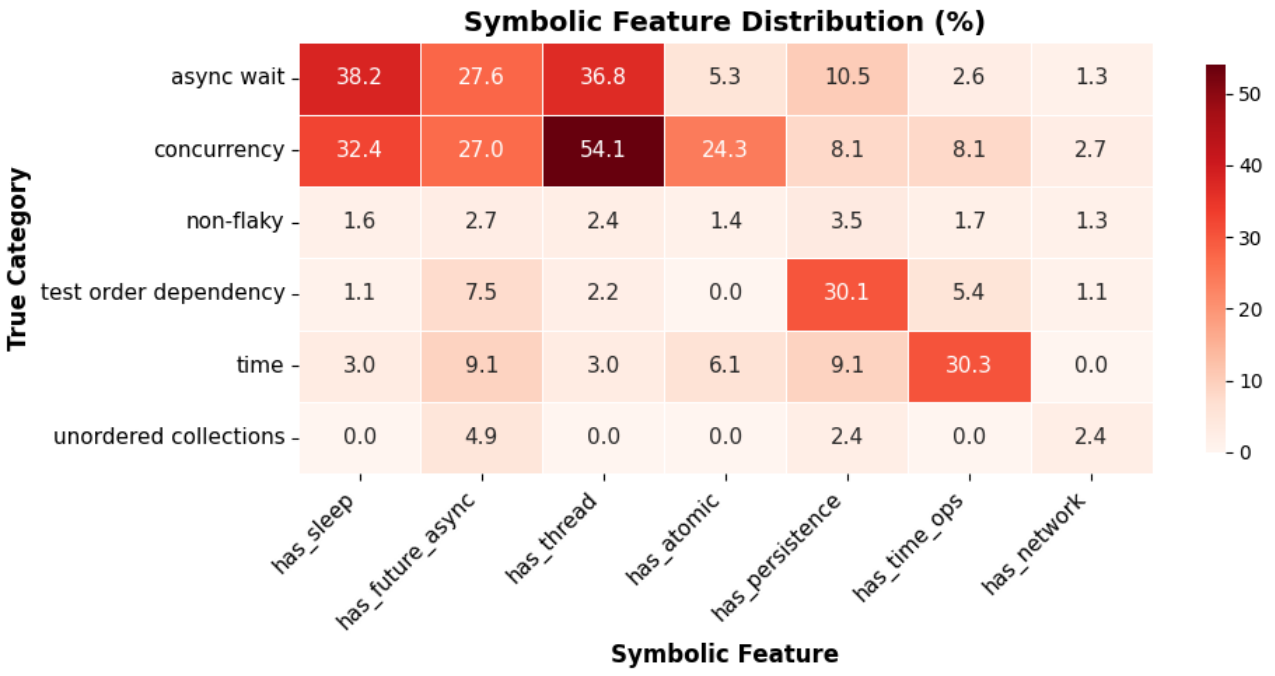

**Figure 3: Distribution of mined token groups across flaky test categories.**

The six flakiness categories evaluated in our study are adopted from the foundational flaky test taxonomy established by Luo et al. [27] and are consistent with the FlakeBench benchmark. Table 1 details the specific "latent" indicators discovered by NeuroFlake.

## 3.2 Neuro-Symbolic Architecture

We utilize GraphCodeBERT [13] as the backbone encoder for fine-tuning and constructing our model. GraphCodeBERT is built on the Transformer architecture and adopts a pre-training objective based on Data Flow Graphs (DFG). This design effectively captures variable dependency and data flow, which is crucial for our problem. We choose GraphCodeBERT over CodeBERT because GraphCodeBERT incorporates semantic structure of code, specifically data flow information, during pre-training. In contrast, CodeBERT [12] mainly relies on the raw sequence of code tokens and associated natural language comments. Therefore, GraphCodeBERT provides a stronger semantic baseline than standard BERT or CodeBERT models used in prior similar tasks.

**Fine-tuning.** We fine-tune the GraphCodeBERT model to distinguish between six target categories. The output of the GraphCodeBERT encoder is a dense vector of length 768, which is the standard, pre-trained hidden dimension size dictated by the foundational GraphCodeBERT and RoBERTa [13, 26] architectures. Directly concatenating this vector with raw symbolic counts leads to a modality mismatch. Therefore, the symbolic vector is fed into a non-linear projection head consisting of two linear layers with ReLU activation and Layer Normalization: $16 \rightarrow 128$. The raw symbolic vector is passed through a sequence of Linear $\rightarrow$ ReLU $\rightarrow$ Linear $\rightarrow$ LayerNorm $\rightarrow$ ReLU layers, mapping it to a 128-dimensional embedding. We then concatenate the pooled semantic vector (768) and the projected symbolic vector (128) to form a fused representation of length 896. This fused vector is passed to two distinct prediction heads: a Binary Head (Flaky vs. Non-Flaky) and a Categorical Head (Specific Flakiness Type). ReLU activation is used, and Dropout layers with a rate of 0.3 are applied during training to prevent overfitting. This strict dropout is critical to avoid shortcut learning, where the model may over-rely on symbolic features. Finally, we use the AdamW optimizer with a learning rate of $2 \times 10^{-5}$ and a weight decay of 0.01.

**Imbalance-Aware Optimization.** A major challenge in our dataset is extreme class imbalance, where Non-Flaky tests significantly outnumber rare categories such as concurrency. To address this, we implement a dual-strategy optimization. First, we compute class weights using the Effective Number of Samples (ENS) method proposed by Cui et al. [8], setting the hyperparameter $\beta = 0.9999$.

The theoretical volume of the data manifold $E_n$, and the resulting loss weight $w_c$ for class $c$, are defined as:

$$E_n = \frac{1 - \beta^n}{1 - \beta} \tag{2}$$

**Table 1: Latent Indicators and Root Causes of Flaky Tests Identified by NeuroFlake**

| Type of Flakiness | Mined Token | Root Cause |
|---|---|---|
| Async Wait | sleep / await / timeunit | Fixed delays cause nondeterministic failures due to varying execution timing; the test waits incorrectly, leading to timeouts or incomplete async events. |
| Async Wait | has_network (Socket, Http, URL, Connect, Netty) | Network latency/failures lead to unpredictable pass/fail without retries. |
| Async Wait | has_future_async (Future, Completable-Future, Promise, CountDownLatch) | Async completion order varies due to thread scheduling, causing unhandled futures. |
| Concurrency | Thread / ExecutorService / Runnable | Multi-threading introduces race conditions and shared state corruption. |
| Concurrency | has_atomic (AtomicInteger, Atomic-Boolean) | Atomic updates fail under contention without locks, leading to visibility issues. |
| Concurrency | has_sync_lock (CountDownLatch, CyclicBarrier, Semaphore, Lock, synchronized) | Incorrect lock timing results in deadlocks or inconsistent access. |
| Time | has_time_ops (currentTimeMillis, nanoTime, Stopwatch, Clock, Duration) | Time assertions fail from clock skew or execution speed variations. |
| Unordered Collections | has_json_unordered (JSON, Map, Set, Iterator) | Order assumptions in collections fail due to JVM hash nondeterminism. |
| Test Order Dependency | has_persistence (.save(), .delete(), Repository, Database, FileSystem) | State leakage (uncleaned files/DBs) causes failures depending on test order. |

$$w_c = \frac{1-\beta}{1-\beta^{n_c}} \tag{3}$$

Unlike prior approaches that use heuristic class weights, ENS assigns higher weights to under-represented classes based on the volume of the data manifold rather than simple inverse frequency.

Second, we replace standard cross-entropy loss with Hybrid Focal Loss for binary classification. The loss is defined as:

$$FL(p_t) = -\alpha(1-p_t)^{\gamma}\log(p_t), \tag{4}$$

where $\gamma = 2.0$ and $\alpha = 0.25$. This focuses gradient updates on hard-to-classify examples and balances precision.

We train for a maximum of 8 epochs with a batch size of 16. Contemporary fine-tuned LLM models typically use 30 epochs, which requires massive hardware resources. We found that 8 epochs are sufficient for convergence due to the high signal-to-noise ratio provided by the neuro-symbolic fusion.

## 4 Experimental Setup

This section describes the experimental design used to evaluate NeuroFlake. We detail the datasets, baseline methods, model configuration, evaluation metrics, and runtime environment to ensure reproducibility and fair comparison.

### 4.1 Dataset

We evaluate our framework on the FlakeBench dataset [40], which consists of 8,574 Java tests from 97 real-world open-source projects. Among 8574 tests, only 280 tests are flaky across five categories (Async Wait, Concurrency, Time, Unordered Collections, Test Order Dependency) and 8294 are non-flaky tests. Prior researches were based on datasets like FlakeFlagger [3] (balanced but artificial, with 1,000 flaky/non-flaky pairs created by rerunning test suites of 24 Java projects 10,000 times.) which contains only binary labels and iDFlakies [18] ( larger but less category-diverse, focusing on order-dependent flakiness). iDFlakies identified 422 flaky tests across 82 projects with a heavy bias toward Order-Dependent tests (50.5%). This limits its diversity regarding other root causes like Concurrency or Async Wait. Although FlakyCat [1] dataset used in the most related work [31], assigns a label to each flaky test based on the category of flakiness root cause, lacks negative samples (relevant non-flaky tests). Flakycat dataset consists of 451 flaky tests across five categories. Models evaluated solely on FlakyCat might have over-estimated prediction accuracy due to potential biases or misrepresentation of real-world test distribution. In real-world software projects, non-flaky tests significantly outnumber flaky ones. FlakeBench mirrors this actual distribution by incorporating a large, high-quality pool of realistic non-flaky tests (8,294 tests).When existing state-of-art models were re-evaluated on FlakeBench, their F1-scores dropped significantly compared to their performance on older benchmarking datasets like FlakyCat. Therefore, we use the FlakeBench dataset which is more robust, recent and captures real world software industry grade imbalance test distribution .

### 4.2 Baselines

To assess the performance of NeuroFlake, we benchmark it against comprehensive suites of frameworks, ranging from black box predictors to state-of-the-art transformer based architectures. We categorize these baselines into three distinct groups.

**1. State-of-the-Art (SOTA) & Specialized Models:** We compare our framework with FlakyLens [40], the current state-of-the-art approach that employs a leakage-aware fine-tuning approach on CodeBERT. Comparing with FlakyLens allows us to measure improvements over the best-performing existing classification task on the FlakeBench dataset. We also evaluate FlakyCat [1], which utilizes Siamese networks and triplet loss for few-shot categorization.Though FlakyCat is built for small data, it evaluates the limits of few-shot learning in larger, messier environments. Furthermore, we include Flakify [11], a remarkable black-box model that relies on code metrics and language-agnostic features rather than semantic understanding. Flakify serves as a baseline for pre-trained LLM methodologies.

**2. Neural & Naive Neuro-Symbolic Baselines:** To isolate the contribution of our specific architectural choices, we implement two controlled variants of LLM fine tuning. First, we evaluate Standard CodeBERT [12] Fine-Tuning, a pure neural baseline that processes the source code without any symbolic augmentation. It is trained on both natural language and programming language, learning connection between them. This quantifies the raw semantic capability of pre-trained models on this task. Standard codebert treats raw source code as a tokenized sequence of texts. Each token is encoded by mapping into a unique ID, and these IDs are furthermore converted into dense vector embeddings. These embeddings are fed into the Transformer model (similar to BERT), that uses the attention masking to understand the relation between tokens, capturing the source code's underlying meaning or semantics. But it misses the latent semantics from code structure because it does not explicitly use Abstract Syntax Trees (ASTs) or data flow. Second, to validate the necessity of our Discriminative Token Mining (DTM), we construct a Naive Neuro-Symbolic Baseline (CodeBERT + Hardcoded). Instead of statistically mining tokens, this model relies on hardcoded symbolic tokens. We do this to observe how some token embeddings might help in flakiness classification tasks compared to pure codebert baseline. This mechanism concatenates the CodeBERT embedding with a vector of manually selected, "common knowledge" flakiness indicators (e.g., Thread.sleep, Future, Random,currentTimeMillis). This comparison is critical to demonstrate that data-driven automated token mining discovery outperforms human-defined heuristics.

**3. Ablation Study (NeuroFlake with/without Symbolic):** Finally, we explicitly quantify the lift provided by our Dual-Channel architecture. To observe the real contribution of our Dual-Channel architecture, we evaluate an ablated version of our own model: NeuroFlake-Neural Only. In this configuration we disable the Adaptive Symbolic Channel and the Late Fusion mechanism and rely solely on the GraphCodeBERT-based Neural Channel. Things to be noted is that, we don't use the graphcodebert pure neural baseline like (section-x-y), rather we use the stronger version of graphcodebert fine-tuning stated in section (methodology). We just omit the symbolic feature embeddings. This only neural channel architecture is similar to our proposed dual channel's neural architecture. To evaluate the explicit contribution of automated mined token embeddings , this ablation setup is necessary. The performance gap between this ablation and the full NeuroFlake framework directly correlates to the signal-to-noise ratio improvement provided by our proposed neuro-symbolic fusion strategy.

## 4.3 NeuroFlake setup

*4.3.1 Data Partitioning: Stratified Project-Wise Disjoint Split.* Prior flaky test classification works rely on random spitting, allowing tests from the same project to appear in both training and test sets. This enables model to memorize project-specific identifiers(e.g., myCompanyConfig).

Hence, we use a Stratified Project-Wise Disjoint Split where tests from the same project never appear in both training and test sets. This guarantees strict separation between seen and unseen repositories and cross project generalization.

1. **Hierarchy-Based Project Categorization:** Standard stratification often fails in multi-label scenarios where a single project contains massive Non-Flaky tests but a few critical Concurrency bugs. We instead employ a Priority-Based Labeling Strategy. We assign each project a primary label corresponding to its most critical flakiness type found, following the hierarchy:

$$\text{Concurrency} > \text{Async} > \text{OD} > \text{Time} > \text{UC} > \text{Non-Flaky}$$

For example, a project with 120 stable tests and 5 Concurrency bugs is labeled as Concurrency. This ensures projects containing rare defects are explicitly tracked.

2. **Stratified Round-Robin Partitioning:** We distribute these labeled projects across $K = 4$ folds using a Stratified Round-Robin approach. Projects within each priority category are shuffled and assigned to folds (fold = i mod K) sequentially. This ensures that every fold receives a balanced distribution of "hard" flakiness types (e.g., Order Dependent) while maintaining strict project-level disjointness (Di_train disjoin Di_test=null)

$$(\mathcal{D}^i_{train} \cap \mathcal{D}^i_{test} = \emptyset).$$

*4.3.2 Model Reloading.* Furthermore, we implement a "Fresh Reload" strategy. At the beginning of each validation fold, the model weights are completely re-initialized to their pre-trained base state. This ensures that the model in Fold $K$ retains zero residual knowledge from Fold $K - 1$. This prevents data leakage and simulates a Zero-Shot scenario where the tool must classify tests from entirely unseen repositories.

## 4.4 Evaluation Metrics

We used Macro F1-score (average=None, zero_division=0) [9, 32], per-class F1 to evaluate classification task, and percentage points (pp) drops under perturbations (rename, deadcode, both) to evaluate the robustness of the augmented model. Results are averaged over F1-score of each category (C), which ensures the minority classes are not dominated by majority classes.

## 4.5 Runtime Environment

We perform fine-tuning and evaluation on Linux machine (Ubuntu 22.04.4 LTS, 64-bit) equipped with AMD Ryzen 5 5500U processor (12 cores with Radon Graphics), 8GB RAM, and an NVIDIA GeForce GTX 1650 GPU (4 GB VRAM), utilizing CUDA 11.8. This is resource-efficient setup, runnable on Google Colab's free T4 tier, as it is using 8 epochs per fold instead of 30 epochs in baselines [40] due to symbolic fusion.

# 5 Evaluation and Discussion

To rigorously assess the effectiveness and robustness of our proposed neuro-symbolic framework, we investigate the following three research questions:

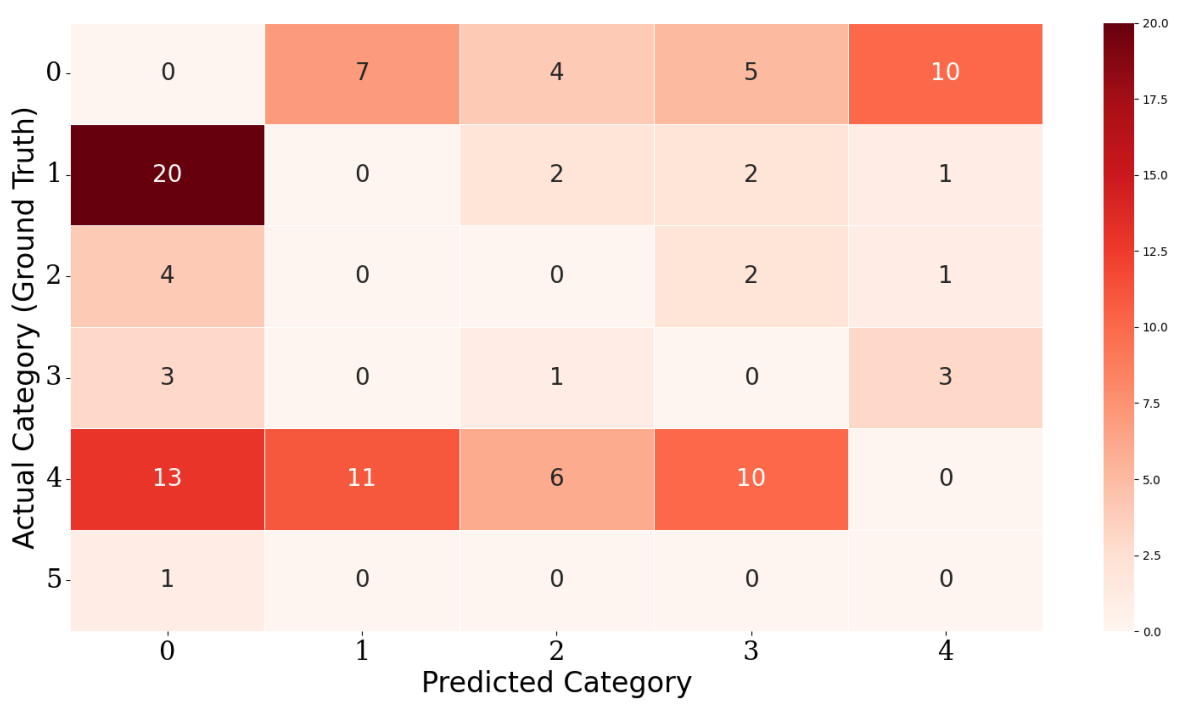

**Figure 4: Confusion matrix for FlakyLens.**

- **RQ1:** How effectively does NeuroFlake classify rare-class flaky tests compared to existing classifiers?
- **RQ2:** What is the contribution of the Adaptive Symbolic Channel to the overall performance?
- **RQ3:** How robust is NeuroFlake against unseen syntactic perturbations (e.g., deadcode, renaming)?

## 5.1 RQ1: NeuroFlake Effectiveness

To quantify how effectively our proposed framework can classify each of the flaky groups we compare NeuroFlake's result with other LLM based test classifiers FlakyFy[11] latest with no data leakage, FlakyCat[1] and FlakyLens[40]. We use FlakeBench dataset for all these three models. The results show each classifier's F1-scores for each category. Categories are Async Wait (Async.), Concurrency (Conc.), Time (Time),Unordered Collections (UC), Test Order Dependency (OD), and non-flaky (Non-flaky). Table 2 shows the performance comparison against SOTA baselines.

We observe that the F-1 score of the asynchronous wait class of our models is increased by 4.57 pp than FlakyLens and 20.14 pp than FlakyFy. While asynchronous wait class is one of the hardest classes due to its similarity with concurrency class. Also it has significant overlaps with the time class. That's why most of the existing models struggle to distinguish between these three classes. Often they miss the actual async test and feed into other classes (Conc, Time). That means much more false negatives for the async class. The most recent model FlakyLens can accurately predict 50 asynchronous flaky tests out of 76 while our model predicts 62 asynchronous flaky tests. This leads to a higher F1-score compared to others.

We further observe that each of the existing SOTA baselines suffer from Conc. class while NeuroFlake still outperforms them for conc class. While existing models have conc class F1 below 40%, NeuroFlake achieves 41.5% for this class. That means 11.89 pp, 21.51 pp and 5.59 pp more than FlakyFy, FlakyCat, FlakyLens respectively. It is due to a small number of samples in conc class. Only 37 conc flaky tests out of total 8574 tests.

If we look at the heatmaps of FlakyLens and NeuroFlake from Figure 4 and Figure 5 (horizontal axis represents predicted category and vertical axis represents actual category) respectively, we notice that FlakyLens's conc class has 18 false positives while NeuroFlake has only 5. That means 13 more concurrency false positives of FlakyLens model leads to poor F1 value.

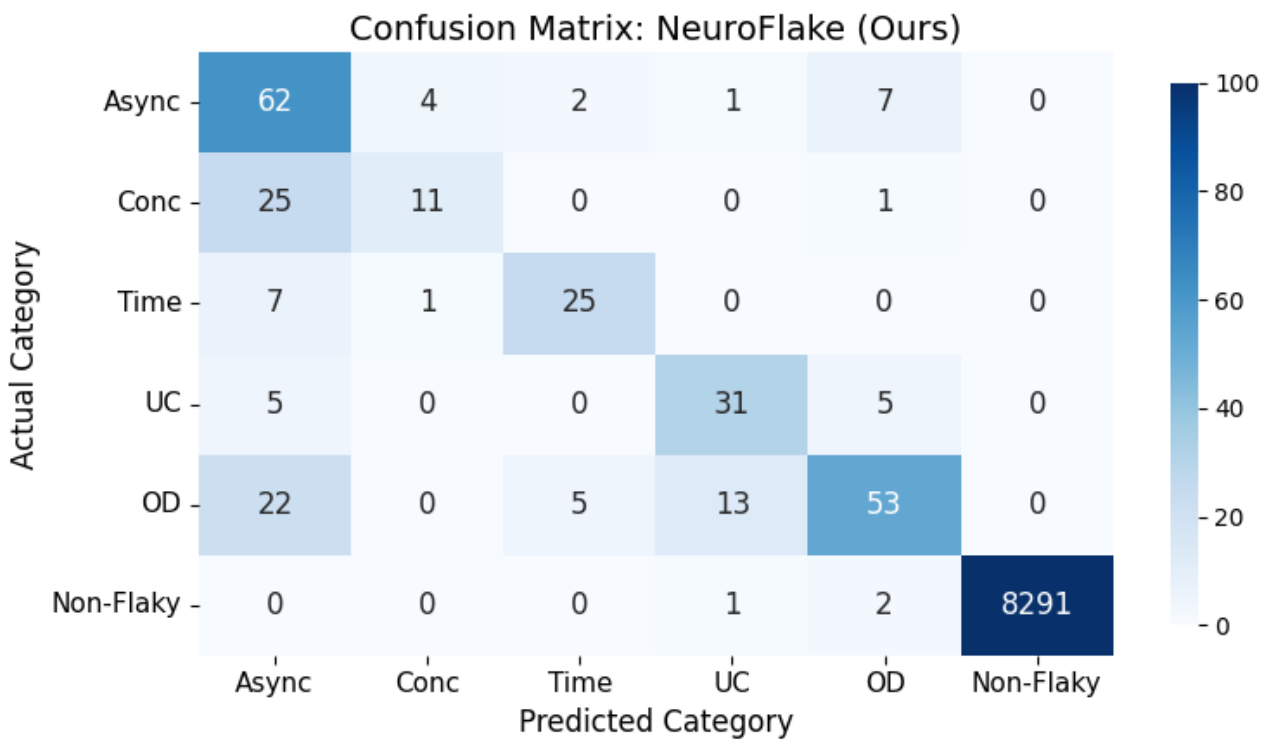

**Figure 5: Confusion matrix for NeuroFlake**

Surprisingly, despite of having smallest sample size, all the models predict the Time class well due to specific signals (e.g., currentTimeMillis, nanoTime). Interestingly, f1 scores of non-flaky class for all models are close to 100 % due to the dominance of non-flaky class.

NeuroFlake outperforms SOTA baselines in 4 of 5 flaky categories. The exception is Unordered Collections (UC), where performance drops. UC flakiness stems from runtime memory states and implicit iteration orders of ubiquitous structures (e.g., `HashSet`, `HashMap`). Because these appear universally in non-flaky code, our Discriminative Token Mining (DTM) struggles to extract distinct symbolic tokens, modestly limiting fusion efficacy for this root cause. Despite this, NeuroFlake's overall macro F1 is 12.72 pp, 17.34 pp, and 3.55 pp higher than FlakyFy, FlakyCat, and FlakyLens, respectively.

That means NeuroFlake outperforms all the SOTA baselines with significant classification improvement. We further analyze the heatmaps in Figure 4 and Figure 5 to achieve a better understanding of misclassified classes. The confusion matrices demonstrate that while both models struggle with the inherent semantic overlap between Async Wait, Concurrency, and OD, our approach demonstrates significantly sharper decision boundaries. A critical weakness in the baseline (FlakyLens) is high false positive rate for Concurrency, which potentially leads developers to debug non-existent race conditions. In contrast, our model successfully distinguishes these categories, achieving perfect separation with zero OD tests misclassified as Concurrency.

Furthermore, while the baseline scatters errors for minority classes like Concurrency across four unrelated classes, our model scatters conc error across two classes. Same for Time and Unordered Collections as well, their misclassified tests are scattered across unrelated categories. Our model's misclassifications are tightly clustered within semantically relevant groups (e.g., confusing Conc, OD only with Async). This ensures that even incorrect predictions remain directionally useful for debugging.

## 5.2 RQ2 : Contribution of the Adaptive Symbolic Channel

Now the question arises whether the Adaptive Symbolic Channel acts as a necessary guide for the model, providing high-level 'semantic anchors' that pure neural embeddings might miss. This evaluation determines if these explicit tokens really help the model to effectively classify the maximum number of flaky tests. To observe the explicit contribution of the adaptive symbolic channel we compared our final model against three internal variants:

(1) **CodeBERT (Pure Neural Base):** The pure neural baseline (fine-tuned microsoft/codebert-base) processes only raw source code without symbolic augmentation. We do not use effective number of samples (ENS) weighting or smart focal loss, instead relying on standard cross-entropy loss. This intentionally simplified setup isolates the predictive capability of pure neural CodeBERT and motivates our choice of GraphCodeBERT as the pretrained backbone.

(2) **CodeBERT + Hardcoded Symbols:** This variant manually maps selected keywords (e.g., `Thread.sleep`, `async`) to symbolic categories without adaptive embeddings. All model parameters and hyperparameters are kept identical to the proposed approach, except that the adaptive symbolic channel is disabled. This setting quantifies the contribution of symbolic features over the pure neural baseline and highlights the limitations of intuition-driven feature selection.

(3) **GraphCodeBERT (No Symbols):** This variant leverages dataflow graphs (DFG) without explicit symbolic category embeddings. All other components remain identical to the final model, with only the adaptive symbolic channel disabled. This configuration directly measures the necessity of adaptive symbolic integration.

**Table 2: F1-score comparison against SOTA baselines. The columns of the table represent different types of flaky test categories (Async.,Conc.,Time,UC,OD) along with non-flaky category and the last column represents average macro F1.The first four rows represent result of different flaky test classification frameworks. The last rows represents the improvement of NeuroFlake's classification task over SOTA baselines.**

| Frameworks | Async. | Conc. | Time | UC | OD | Non-flaky | Macro Avg. |
|---|---|---|---|---|---|---|---|
| Flakify | 42.80 | 29.62 | 72.55 | 39.15 | 54.70 | 100.00 | 56.62 |
| FlakyCat | 43.75 | 20.00 | 53.75 | 58.75 | 36.25 | 99.50 | 52.00 |
| FlakyLens | 58.37 | 35.92 | 72.73 | 73.63 | 64.35 | 100.00 | 65.79 |
| NeuroFlake (ours) | 62.94 | 41.51 | 76.92 | 69.70 | 65.80 | 99.98 | 69.34 |
| ↑ over FlakyLens | 4.57 | 5.59 | 4.19 | -3.93 | 1.45 | -0.02 | 3.55 |

Table 3 shows the comparison between these three variants.

We observe that the f1 score of pure neural codebert is significantly low. The "Raw" CodeBERT serves as a weak baseline, confirming that neural language models alone struggle to disentangle complex flakiness causes like Concurrency or Time-dependency. The hardcoded symbolic codebert gains a massive improvement compared to the pure one. Hardcoding symbols improves performance but is limited by the rigidity of manual rules. This implies the impact of the adaptive symbolic channel. Finally the GraphCodeBERT (No Symbols) model shows a surprising result we can observe from the table 3. It achieves f1 score 61.35% which is even better than the hardcoded smarter codebert which achieves f1 score 60.44%. This analysis justifies the reason behind choosing GraphCodeBERT over codebert as base model. Graphcodebert alone can predict the flaky classes better than the codebert with hardcoded symbols. It's because codebert treats the code as a sequence of plain texts while GraphCodeBERT captures the internal graph structure and true semantics. So we leverage the capacity of GraphCodeBERT along with The adaptive symbolic channel and get the best performing model with the highest f1 score.

## 5.3 RQ3: NeuroFlake Robustness

Neural models, including Large language models (LLMs) for code often learn shortcut patterns—superficial textual artifacts (e.g., specific variable names like `waitTime` or method calls like `Thread.sleep`)—instead of understanding the underlying semantic logic. As a result, adversarial yet semantics-preserving transformations can significantly degrade performance. We collectively refer to these transformations as perturbations, defined as syntax- or text-level modifications that do not change execution behavior. This phenomenon of LLM is known as shortcut learning [46]. Shortcut learning may improve efficiency in LLMs [30] but makes them vulnerable to adversarial perturbations which leads to misclassification. For example , in Figure 6 we observe that, code at the left side (a) is a true non-flaky test. But after injecting deadcodes (b), it looks like async. wait flaky tets through which LLM is tricked even the semantic meaning is same. A reverse case is observed in Figure 7, where (a) represents an original Time flaky test and (b) represents the renaming perturbed code. We then convert the meaningful variable names to arbitrary names which causes true flaky tests to be predicted as non-flaky.

Prior work [40] reported that FlakyLens (fine-tuned CodeBERT) suffers a severe drop of 18.37 pp and 10.98 pp in f1 score for deadcode and perturbation variables. To mitigate this brittleness, we introduce targeted augmentation during training. This augmentation strategy explicitly teaches the model to ignore deceptive tokens and focus on semantic structure. Importantly augmentations are applied to both flaky and non-flaky tests (with higher probability 0.95 for rare flaky categories) because non-flaky tests can also be falsely predicted as flaky under perturbations.

*5.3.1 Adversarial Augmentation Design.* It consists of three complementary components:

(1) **Aggressive Variable Obfuscation** User-defined variables are renamed (e.g., StartTime ⟶ VAR_3) to ensure the model does not memorize specific identifiers.

(2) **Adversarial Structural Noise Inserts Unreachable Blocks Containing Decoy Tokens** Inserts dead codes intentionally

**Table 3: F1 score comparison between three internal veriants (Pure CodeBERT,CodeBERT with hardcoded signal embeddings, GraphCodeBERT without symbolic channel) of NeuroFlake to analyze the impact of adaptive symbolic channel**

| Internal Variants of Poposed Framework | Async. | Conc. | Time | UC | OD | Non-flaky | Macro Avg. |
|---|---|---|---|---|---|---|---|
| CodeBERT | 24.00 | 0.00 | 0.00 | 0.00 | 0.00 | 100.00 | 20.70 |
| CodeBERT + Hardcoded Symbols | 63.04 | 48.39 | 71.19 | 46.72 | 33.33 | 99.95 | 60.44 |
| GraphCodeBERT (No Symbols) | 59.22 | 35.47 | 71.02 | 61.25 | 41.17 | 99.96 | 61.35 |

**(a) Original: Non-Flaky Test**

```
@Test
public void serializationRoundTripTest() throws Exception {
    // Test serializing and deserializing
    final Event event = Event.create("create", "foo", "nginx",
        Event.Type.CONTAINER,
        Event.Actor.create("bar",
            ImmutableMap.of("image", "nginx", "name", "dckr")),
        new Date(1487356000), 100L);

    final ObjectMapper mapper = ObjectMapperProvider.mapper();
    final String json = mapper.writeValueAsString(event);
    final Event event2 = mapper.readValue(json, Event.class);
    assertThat(event, equalTo(event2));
}
```

**(b) Predicted: Async Wait (Deadcode)**

```
@Test
public void serializationRoundTripTest() throws Exception {
    // [TRAP] Deadcode with "Async" signals:
    while (false) {
        long timeout = 5000;
        while (timeout > 0) {
            Thread.sleep(100); // Decoy Token
            timeout -= 100;
            if (Event.create("dummy") != null) break;
        }
        CountDownLatch latch = new CountDownLatch(1);
        latch.await(); // Strong Async Signal
    }

    // Actual logic (Same as original)
    final Event event = Event.create("create", "foo", ...);
    // ... remainder of original code ...
}
```

**Figure 6: Adversarial Perturbation. In (b), we inject unreachable code containing high-attention tokens (orange). The logic remains Non-Flaky, but the baseline model is tricked.**

containing deceiving tokens—highly predictive concurrency-related keywords (e.g., `Thread.sleep`, `CompletableFuture`)—but only in:

- `while(false)` or equivalent unreachable blocks,
- exception handlers that never execute,
- string literals or comments.

(3) **Decoy Token Injection** Deceptive tokens are placed inside comments and print statements so that the model learns to distinguish code semantics from textual artifacts.

During training, augmentations are applied with a 50% probability. This ensures the model observes both clean and perturbed code. This avoids overfitting to adversarial patterns only but still ensures robustness. This introduces a trade-off : effectiveness vs robustness.

- Without augmentation: clean test macro-F1 = 69.34%
- With augmentation: clean test macro-F1 = 64.84%

**(a) Original: Time Flaky (Semantic Variables)**

```
@Test
public void testMatchesSpeedTest() throws Exception {
    int iterations = 15;
    String password = new RandomValueStringGenerator().generate();
    String encodedBcrypt = cachingPasswordEncoder.encode(password)
     ;

    long nanoStart = System.nanoTime();
    for (int i = 0; i < iterations; i++) {
        assertTrue(cachingPasswordEncoder.getPasswordEncoder()
            .matches(password, encodedBcrypt));

        long nanoStop = System.nanoTime();
        long bcryptTime = nanoStop - nanoStart;

        nanoStart = System.nanoTime();
        for (int j = 0; j < iterations; j++) {
            nanoStop = System.nanoTime();
            long cacheTime = nanoStop - nanoStart;
            // Assertion relies on semantic time variables
            assertTrue(bcryptTime > (10 * cacheTime));
        }
    }
}
```

**(b) Predicted: Non-Flaky (Renamed)**

```
@Test
public void testMatchesSpeedTest() throws Exception {
    int _loopMax = 15; // Was "iterations"
    String _s1 = new RandomValueStringGenerator().generate();
    String _s2 = cachingPasswordEncoder.encode( _s1 );

    // Renaming Attack: Stripping "Time" semantics
    long _t1 = System.nanoTime();
    for (int i = 0; i < _loopMax ; i++) {
        assertTrue(cachingPasswordEncoder.getPasswordEncoder()
            .matches( _s1 , _s2 ));

        long _t2 = System.nanoTime();
        long _valA = _t2 - _t1 ; // "bcryptTime"

        _t1 = System.nanoTime();
        for (int j = 0; j < _loopMax ; j++) {
            _t2 = System.nanoTime();
            long _valB = _t2 - _t1 ; // "cacheTime"
            // Model fails to link _valA/_valB to time
            assertTrue( _valA > (10 * _valB ));
        }
    }
}
```

**Figure 7: Semantic Masking Example. In (b), we rename time-specific variables (e.g., `nanoStart` → `_t1`). While logic remains Time Flaky, the removal of semantics confuses the baseline.**

But yet the f1 drop after applying augmentation is not that major compared to out NeuroFlake baseline . This augmented version sacrifices 4.50 pp of f1 scores which ensures a controlled robustness–effectiveness trade-off, consistent with adversarial training literature.

*5.3.2 Stress Testing Under Unseen Perturbations.* To fairly evaluate robustness, we perform stress tests using unseen perturbations. Here unseen means distinct from those observed during training. This avoids the pitfall of "memorized defenses." Example : Training augmentation: `if(false) { Thread.sleep(100); }` Test perturbation: `while(false) { Thread.sleep(100); }`

Although syntactically different, they are semantically equivalent. Table 4(Section A) shows the comparison of f1 scorers among clean tests vs unseen perturbed tests of NeuroFlake's augmented version and drops in each perturbed version. Under these unseen attacks, our augmented model experiences only a 4-6% macro-F1 drop, substantially lower than the 8-18.37% drop reported in prior work[40]. Moreover, compared to FlakyLens[40], NeuroFlake exhibits achieving 11.76% and 6.72% lower performance drops under deadcode and variable renaming perturbations, respectively. These results demonstrate that the proposed augmentation strategy improves generalization to unseen perturbations rather than overfitting to specific syntactic patterns.

*5.3.3 Role of the Neuro-Symbolic Design in Robustness Testing.* To isolate the contribution of the Adaptive Symbolic Channel, we conduct an ablation study by disabling discriminative token mining and using fixed symbolic features. The motivation is to find out whether the adaptive symbolic channel helps in achieving higher f1 under perturbation or not.

Results show:

- Clean test macro-F1 drops further to 60.26.
- Perturbation drops remain small (≈4–6%).

Table 4 (Section B) depicts the F1-score performance under adversarial perturbations without the adaptive symbolic channel.

This indicates that symbolic features contribute not only to accuracy but also to robustness, complementing the semantic representations learned by GraphCodeBERT. Although the average drops in both cases (augmentation with neuro-symbolic architecture, augmentation after disabling the symbolic channel) remain stable (4-7%) which indicates our proposed framework is robust enough to predict flaky categories under perturbation. This mitigates the short-cut learning of LLMs.

**Table 4: Robustness Stress Test. Comparison of F1 scores and performance drops (%) under unseen perturbations.**

| | Orig. | Rename | | Deadcode | | Both | |
|---|---|---|---|---|---|---|---|
| **Cat.** | **F1** | **F1** | **Drop** | **F1** | **Drop** | **F1** | **Drop** |
| ***A. NeuroFlake (With Adaptive Symbolic Channel)*** | | | | | | | |
| Async. | 60.11 | 58.62 | 1.49 | 57.78 | 2.33 | 55.87 | 4.24 |
| Conc. | 29.09 | 31.58 | -2.49 | 32.14 | 3.05 | 30.99 | -1.90 |
| Time | 66.67 | 61.30 | 5.37 | 54.85 | 11.82 | 51.72 | 14.95 |
| UC | 65.82 | 56.10 | 9.29 | 51.28 | 14.54 | 48.48 | 17.34 |
| OD | 67.39 | 55.51 | 11.88 | 59.54 | 7.85 | 53.41 | 13.98 |
| Non-F | 99.97 | 99.96 | 0.01 | 99.92 | 0.07 | 99.90 | 0.09 |
| *Avg Drop* | – | – | **4.26** | – | **6.61** | – | **8.11** |
| ***B. Ablation (No Symbolic Channel, fixed features only)*** | | | | | | | |
| Async. | 57.13 | 46.94 | 10.19 | 56.23 | 0.90 | 41.39 | 15.74 |
| Conc. | 31.80 | 32.41 | -0.61 | 33.10 | -1.30 | 37.74 | -5.94 |
| Time | 57.95 | 55.50 | 2.45 | 50.05 | 7.90 | 46.45 | 11.50 |
| UC | 63.31 | 54.19 | 9.12 | 56.30 | 7.01 | 44.34 | 18.97 |
| OD | 51.37 | 46.58 | 4.79 | 35.60 | 15.77 | 26.97 | 24.40 |
| Non-F | 99.99 | 99.97 | 0.02 | 99.80 | 0.19 | 99.86 | 0.13 |
| *Avg Drop* | – | – | **4.32** | – | **5.08** | – | **10.08** |

## 6 Threats to Validity

## 6.1 Internal Validity

Our study relies on FlakeBench [40]. While we assume the ground truth labels are accurate, potential noise in the original dataset could affect training outcomes. To mitigate label noise, we excluded ambiguous tests. A primary internal threat relates to our Discriminative Token Mining (DTM): it identifies tokens based on statistical correlation, which does not guarantee a strict causal relationship with the underlying root causes of flakiness. Furthermore, while our cross-project consistency threshold ($N \geq 3$) suppresses project-specific noise, it may inadvertently over-filter long-tail, framework-specific signals that are diagnostically valuable for minority classes. Finally, our adversarial augmentations (e.g., dead-code injection) act as strict stress tests against shortcut learning, but they may diverge from realistic developer refactoring patterns.

## 6.2 External Validity

We evaluate NeuroFlake primarily on Java projects; projects written in other languages (e.g., Python) may exhibit different flakiness patterns. Future work is required to validate NeuroFlake across multi-language corpora.

## 6.3 Construct Validity

We assess classification performance using macro F1-score and per-class F1-scores. While these metrics are standard for imbalanced classification, they may not fully capture the industrial cost trade-offs between false alarms and missed flaky tests in real-world CI pipelines. We partially mitigate this limitation by analyzing relative performance degradation under unseen perturbation settings rather than relying solely on absolute scores.

## 7 Conclusion

We popose NeuroFlake, a neuro-symbolic framework for robust flaky test classification. Existing LLM based approaches suffer from short-cut learning and struggle with severe class imbalance. NeuroFlake overcomes these limitations by integrating semantic code embeddings with structural symbolic features. Also employs a smart adversarial augmentation strategy. Our evaluation on FlakeBench dataset demonstrate that NeuroFlake significantly outperforms state-of-the-art baselines in both classification accuracy and robustness against deceptive code perturbations.

In future, we plan to extend NeuroFlake to support dynamically typed languages such as Python and JavaScript, where flakiness manifests differently and enhance the symbolic feature embeddings by adding dynamic execution traces.

## References

[1] Amal Akli, Guillaume Haben, Sarra Habchi, Mike Papadakis, and Yves Le Traon. 2023. FlakyCat: Predicting flaky tests categories using few-shot learning. In *2023 IEEE/ACM International Conference on Automation of Software Test (AST)*. IEEE, 140–151.

[2] Nauman Bin Ali, Emelie Engström, Masoumeh Taromirad, Mohammad Reza Mousavi, Nasir Mehmood Minhas, Daniel Helgesson, Sebastian Kunze, and Mahsa Varshosaz. 2019. On the search for industry-relevant regression testing research. *Empirical Software Engineering* 24, 4 (2019), 2020–2055.

[3] Abdulrahman Alshammari, Christopher Morris, Michael Hilton, and Jonathan Bell. 2021. Flakeflagger: Predicting flakiness without rerunning tests. In *2021 IEEE/ACM 43rd International Conference on Software Engineering (ICSE)*. IEEE, 1572–1584.

[4] Jonathan Bell, Owolabi Legunsen, Michael Hilton, Lamyaa Eloussi, Tifany Yung, and Darko Marinov. 2018. DeFlaker: Automatically detecting flaky tests. In *Proceedings of the 40th international conference on software engineering*. 433–444.

[5] Nghi DQ Bui, Yijun Yu, and Lingxiao Jiang. 2019. Autofocus: interpreting attention-based neural networks by code perturbation. In *2019 34th IEEE/ACM International Conference on Automated Software Engineering (ASE)*. IEEE, 38–41.

[6] Junkai Chen, Li Zhenhao, Hu Xing, and Xia Xin. 2024. Nlperturbator: Studying the robustness of code llms to natural language variations. *ACM Transactions on Software Engineering and Methodology* (2024).

[7] Yang Chen and Reyhaneh Jabbarvand. 2024. Neurosymbolic repair of test flakiness. In *Proceedings of the 33rd ACM SIGSOFT International Symposium on Software Testing and Analysis*. 1402–1414.

[8] Yin Cui, Menglin Jia, Tsung-Yi Lin, Yang Song, and Serge Belongie. 2019. Class-balanced loss based on effective number of samples. In *Proceedings of the IEEE/CVF conference on computer vision and pattern recognition*. 9268–9277.

[9] Hercules Dalianis. 2018. Evaluation metrics and evaluation. In *Clinical Text Mining: secondary use of electronic patient records*. Springer, 45–53.

[10] Sakina Fatima. 2025. *Detection, Categorization and Repair of Flaky Tests Using Large Language Models*. Ph. D. Dissertation. Université d'Ottawa/University of Ottawa.

[11] Sakina Fatima, Taher A Ghaleb, and Lionel Briand. 2022. Flakify: A black-box, language model-based predictor for flaky tests. *IEEE Transactions on Software Engineering* 49, 4 (2022), 1912–1927.

[12] Z Feng. 2020. Codebert: A pre-trained model for program-ming and natural languages. *arXiv preprint arXiv:2002.08155* (2020).

[13] Daya Guo, Shuo Ren, Shuai Lu, Zhangyin Feng, Duyu Tang, Shujie Liu, Long Zhou, Nan Duan, Alexey Svyatkovskiy, Shengyu Fu, et al. 2020. Graphcodebert: Pre-training code representations with data flow. *arXiv preprint arXiv:2009.08366* (2020).

[14] Negar Hashemi, Amjed Tahir, Shawn Rasheed, August Shi, and Rachel Blagojevic. 2025. Detecting and evaluating order-dependent flaky tests in javascript. In *2025 IEEE Conference on Software Testing, Verification and Validation (ICST)*. IEEE, 13–24.

[15] Pascal Hitzler, Aaron Eberhart, Monireh Ebrahimi, Md Kamruzzaman Sarker, and Lu Zhou. 2022. Neuro-symbolic approaches in artificial intelligence. *National Science Review* 9, 6 (2022), nwac035.

[16] Imen Jaoua, Oussama Ben Sghaier, and Houari Sahraoui. 2025. Combining Large Language Models with Static Analyzers for Code Review Generation. In *2025 IEEE/ACM 22nd International Conference on Mining Software Repositories (MSR)*. IEEE, 174–186.

[17] Wing Lam, Patrice Godefroid, Suman Nath, Anirudh Santhiar, and Suresh Thummalapenta. 2019. Root causing flaky tests in a large-scale industrial setting. In *Proceedings of the 28th ACM SIGSOFT International Symposium on Software Testing and Analysis*. 101–111.

[18] Wing Lam, Reed Oei, August Shi, Darko Marinov, and Tao Xie. 2019. iDFlakies: A framework for detecting and partially classifying flaky tests. In *2019 12th ieee conference on software testing, validation and verification (icst)*. IEEE, 312–322.

[19] Tanakorn Leesatapornwongsa, Xiang Ren, and Suman Nath. 2022. FlakeRepro: Automated and efficient reproduction of concurrency-related flaky tests. In *Proceedings of the 30th ACM Joint European Software Engineering Conference and Symposium on the Foundations of Software Engineering*. 1509–1520.

[20] Fabian Leinen, Daniel Elsner, Alexander Pretschner, Andreas Stahlbauer, Michael Sailer, and Elmar Jürgens. 2024. Cost of flaky tests in continuous integration: An industrial case study. In *2024 IEEE Conference on Software Testing, Verification and Validation (ICST)*. IEEE, 329–340.

[21] Nate Levin, Chengpeng Li, Yule Zhang, August Shi, and Wing Lam. 2025. Takuan: Using Dynamic Invariants to Debug Order-Dependent Flaky Tests. In *2025 IEEE/ACM 47th International Conference on Software Engineering: New Ideas and Emerging Results (ICSE-NIER)*. IEEE, 81–85.

[22] Chengpeng Li and August Shi. 2022. Evolution-aware detection of order-dependent flaky tests. In *Proceedings of the 31st ACM SIGSOFT International Symposium on Software Testing and Analysis*. 114–125.

[23] Haonan Li, Yu Hao, Yizhuo Zhai, and Zhiyun Qian. 2023. Assisting static analysis with large language models: A chatgpt experiment. In *Proceedings of the 31st ACM Joint European Software Engineering Conference and Symposium on the Foundations of Software Engineering*. 2107–2111.

[24] Shizhe Lin, Ryan Zheng He Liu, and Ladan Tahvildari. 2024. FlaKat: A Machine Learning-Based Categorization Framework for Flaky Tests. *arXiv preprint arXiv:2403.01003* (2024).

[25] Xinyue Liu, Zihe Song, Weike Fang, Wei Yang, and Weihang Wang. 2024. Wefix: Intelligent automatic generation of explicit waits for efficient web end-to-end flaky tests. In *Proceedings of the ACM Web Conference 2024*. 3043–3052.

[26] Yinhan Liu, Myle Ott, Naman Goyal, Jingfei Du, Mandar Joshi, Danqi Chen, Omer Levy, Mike Lewis, Luke Zettlemoyer, and Veselin Stoyanov. 2019. Roberta: A robustly optimized bert pretraining approach. *arXiv preprint arXiv:1907.11692* (2019).

[27] Qingzhou Luo, Farah Hariri, Lamyaa Eloussi, and Darko Marinov. 2014. An empirical analysis of flaky tests. In *Proceedings of the 22nd ACM SIGSOFT international symposium on foundations of software engineering*. 643–653.

[28] Wei Ma, Shangqing Liu, Zhihao Lin, Wenhan Wang, Qiang Hu, Ye Liu, Cen Zhang, Liming Nie, Li Li, and Yang Liu. 2023. Lms: Understanding code syntax and semantics for code analysis. *arXiv preprint arXiv:2305.12138* (2023).

[29] Christopher D Manning. 2008. *Introduction to information retrieval*. Syngress Publishing,.

[30] Amane Meibuki, Renshu Nanao, and Mugen Outa. 2024. Improving learning efficiency in large language models through shortcut learning. (2024).

[31] Riddhi More and Jeremy S Bradbury. 2025. An Analysis of LLM Fine-Tuning and Few-Shot Learning for Flaky Test Detection and Classification. In *2025 IEEE Conference on Software Testing, Verification and Validation (ICST)*. IEEE, 349–359.

[32] Gireen Naidu, Tranos Zuva, and Elias Mmbongeni Sibanda. 2023. A review of evaluation metrics in machine learning algorithms. In *Computer science on-line conference*. Springer, 15–25.

[33] Daye Nam, Andrew Macvean, Vincent Hellendoorn, Bogdan Vasilescu, and Brad Myers. 2024. Using an llm to help with code understanding. In *Proceedings of the IEEE/ACM 46th International Conference on Software Engineering*. 1–13.

[34] Owain Parry, Gregory M Kapfhammer, Michael Hilton, and Phil McMinn. 2021. A survey of flaky tests. *ACM Transactions on Software Engineering and Methodology (TOSEM)* 31, 1 (2021), 1–74.

[35] Owain Parry, Gregory M Kapfhammer, Michael Hilton, and Phil McMinn. 2023. Empirically evaluating flaky test detection techniques combining test case rerunning and machine learning models. *Empirical Software Engineering* 28, 3 (2023), 72.

[36] Yu Pei, Sarra Habchi, Renaud Rwemalika, Jeongju Sohn, and Mike Papadakis. 2022. An empirical study of async wait flakiness in front-end testing.. In *BENEVOL*.

[37] Yu Pei, Jeongju Sohn, Sarra Habchi, and Mike Papadakis. 2025. Non-flaky and nearly optimal time-based treatment of asynchronous wait web tests. *ACM Transactions on Software Engineering and Methodology* 34, 2 (2025), 1–29.

[38] Shanto Rahman, Abdelrahman Baz, Sasa Misailovic, and August Shi. 2024. Quantizing large-language models for predicting flaky tests. In *2024 IEEE Conference on Software Testing, Verification and Validation (ICST)*. IEEE, 93–104.

[39] Shanto Rahman, Bala Naren Chanumolu, Suzzana Rafi, August Shi, and Wing Lam. 2025. Ranking Relevant Tests for Order-Dependent Flaky Tests. In *2025 IEEE/ACM 47th International Conference on Software Engineering (ICSE)*. IEEE Computer Society, 715–715.

[40] Shanto Rahman, Saikat Dutta, and August Shi. 2025. Understanding and Improving Flaky Test Classification. *Proceedings of the ACM on Programming Languages* 9, OOPSLA2 (2025), 1345–1371.

[41] Shanto Rahman and August Shi. 2024. FlakeSync: Automatically repairing async flaky tests. In *Proceedings of the IEEE/ACM 46th International Conference on Software Engineering*. 1–12.

[42] Denini Silva, Leopoldo Teixeira, and Marcelo d'Amorim. 2020. Shake it! detecting flaky tests caused by concurrency with shaker. In *2020 IEEE International Conference on Software Maintenance and Evolution (ICSME)*. IEEE, 301–311.

[43] Jiaguo Wang, Yan Lei, Maojin Li, Guanyu Ren, Huan Xie, Shifeng Jin, Junchao Li, and Jian Hu. 2024. Flakyrank: Predicting Flaky Tests Using Augmented Learning to Rank. In *2024 IEEE International Conference on Software Analysis, Evolution and Reengineering (SANER)*. IEEE, 872–883.

[44] Yue Wang, Hung Le, Akhilesh Gotmare, Nghi Bui, Junnan Li, and Steven Hoi. 2023. Codet5+: Open code large language models for code understanding and generation. In *Proceedings of the 2023 conference on empirical methods in natural language processing*. 1069–1088.

[45] Shin Yoo and Mark Harman. 2012. Regression testing minimization, selection and prioritization: a survey. *Software testing, verification and reliability* 22, 2 (2012), 67–120.

[46] Yu Yuan, Lili Zhao, Kai Zhang, Guangting Zheng, and Qi Liu. 2024. Do llms overcome shortcut learning? an evaluation of shortcut challenges in large language models. *arXiv preprint arXiv:2410.13343* (2024).